\documentclass[12pt]{iopart}

\usepackage{graphicx}

\usepackage{iopams}  
\newtheorem{theorem}{Theorem}

\newtheorem{corollary}{Corollary}
\usepackage{color}

\begin{document}

\title[Bifurcation analysis of figure-eight choreography in TBP]
{Bifurcation analysis of figure-eight choreography in the three-body problem
based on crystallographic point groups
}

\author{Hiroshi Fukuda$^1$ and Hiroshi Ozaki$^2$}

\address{$^1$College of liberal arts and sciences, Kitasato University, Minami-ku 1-15-1, Sagamihara, Kanagawa 252-0373, Japan}
\address{$^2$STEM Education Center, Tokai University, 4-1-1, Kitakaname, Hiratsuka, Kanagawa 259-1292, Japan}
\ead{$^1$fukuda@kitasato-u.ac.jp, $^2$ozaki@tokai.ac.jp}
\vspace{10pt}

\begin{abstract}
The bifurcation of figure-eight choreography is analyzed  
by its symmetry group based on the variational principle of the action.
The irreducible representations determine the symmetry and the dimension of the 
Lyapunov-Schmidt reduced action, which yields  
four types of bifurcations in the sequence of the bifurcation cascade. 
Type 1 bifurcation, represented by trivial representation, 
bifurcates two solutions.  
Type 2, by non-trivial one-dimensional representation, 
bifurcates two congruent solutions.  
Type 3 and 4, by two-dimensional irreducible representations,  
bifurcate two sets of three and six congruent solutions, respectively. 
We analyze numerical bifurcation solutions previously published and four new ones: 
non-symmetric choreographic solution of type 2,   
non-planar solution of type 2, $y$-axis symmetric solution of type 3,  
and non-symmetric solution of type 4.  
\end{abstract}

%
\vspace{2pc}

\section{Introduction}\label{sec:intro}
The figure-eight choreography in the three-body problem is 
a motion of three equal masses, chasing each other in the common 
figure-eight-shaped orbit \cite{moore,Chenciner}.
For inhomogeneous interaction potential, the shape of the orbit varies  
as a function of the period of the motion,    
and it bifurcates \cite{sbano,sbano2010,fukuda2017,fukuda2019}.
For homogeneous interaction potential, 
it remains similar independent of the period, while 
bifurcations occur by the power of homogeneous potential or one of the masses of the bodies 
as a parameter \cite{fukuda2019,galan2002,doedel2003}.

The Sim\'{o}'s H solution \cite{simoH} is a famous bifurcation solution from the figure-eight choreography under Newtonian potential. 
It is almost choreographic but has slightly different three figure-eight-shaped orbits.
Such breaking of the choreographic orbit is a sign of the bifurcation.  
This strange solution motivated our study \cite{fukuda2019}. 
Moreover, for the system under inhomogeneous interaction potential,     
the study of the bifurcation is essentially important 
since its period can be a bifurcation parameter 
which can vary without changing the system parameters 
such as masses or the power of the potential.

The figure-eight choreography is a very symmetric motion and actually 
has four symmetry elements forming a crystallographic point group \cite{Chenciner2}. 
The bifurcation of symmetric objects 
is called equivariant bifurcation \cite{CL,IJ}.
In a mathematical paper on equivariant bifurcation of the planar three-body problem \cite{barutello}, 
though the figure-eight choreography is discussed, 
no numerical calculation of its bifurcation is contained.
On the other hand, a paper on numerical calculation of bifurcation \cite{fukuda2019,galan2002,doedel2003} 
does not discuss the symmetry of the bifurcation solutions.   

In this paper, we formulate an equivariant bifurcation analysis of the choreography  
based on the variational principle of the action 
utilizing the second variation of the action \cite{fujiwara2020}. 
It is because we started  
our calculation with the system under Lennard-Jones potential and 
found unexpectedly various choreographies \cite{fukuda2017},  
then analyzed them by Morse index, the number of directions 
that the second derivative of the action is negative at the critical point \cite{fukuda2018}.

In section \ref{sec:nes}, a necessary condition for the bifurcation is 
shown by eigenvalues of an operator related to the second derivative of the action 
and the bifurcation solutions are expressed by its eigenfunctions.
The number of variables to express the bifurcation solution is reduced to 
that of the degeneracy of the eigenvalue, then its action, 
called the Lyapunov-Schmidt reduced action, is also expressed by the same variables. 
In section \ref{sec:symsol}, a theorem and a corollary on its symmetries are introduced 
through the real orthogonal representation of its symmetry group. 
In section \ref{sec:symact}, another corollary on symmetries of the Lyapunov-Schmidt 
reduced action is introduced. 
From the corollaries and the irreducible representation, 
it is shown that four types of bifurcation are possible. 
In section \ref{sec:num}, numerically calculated bifurcation solutions in \cite{fukuda2019} 
and four new solutions are analyzed.
Section \ref{sec:sum} is the summary and discussions.

\section{Necessary condition and representation for bifurcation}\label{sec:nes}
We consider a periodic solution $q(t)=q(t+T)$ with a period $T$ 
of the Euler-Lagrange equations 
\begin{equation}\label{eq.motion}
	\frac{d}{dt} \frac{\partial L}{\partial \dot{q}} = \frac{\partial L}{\partial q}
\end{equation}
for the Lagrangian of the equal mass three-body problem with interaction potential $u$:
\begin{equation}\label{eq.L}
	L = \frac{1}{2}\dot{q}^2-U, \;
	U = \sum_{i>j}u(|\bi{r}_i-\bi{r}_j|),	  
\end{equation}
where $\bi{r}_i=(x_i,y_i,z_i)$, $i=1,2,3$ 
are the position vectors of body $i$ and 
\begin{equation}
	q=(q_1,q_2,\ldots,q_N)=(x_1,y_1,z_1,x_2,y_2,z_2,x_3,y_3,z_3),
\end{equation}
with $N=9$.
We use matrix notation, while we do not distinguish column and row vectors strictly 
if it is obvious from the context.

\subsection{Necessary condition for bifurcation}
Suppose a periodic solution $q$ depends on a parameter $\xi$, 
such as the period $T$ or the order of homogeneous potential 
and bifurcates into $q^{(b)}$ at $\xi=\xi^{(b)}$.
Since both $q$ and $q^{(b)}$ are solutions of (\ref{eq.motion}) with (\ref{eq.L}), 
their difference
\begin{equation}\label{Deltaq}
	\Delta q = q^{(b)}-q
\end{equation} 
satisfies a relation made of the Hessian of the Lagrangian 
\begin{equation}\label{hessian}
	\frac{d^2}{dt^2}\Delta q 
	=
	\left.\frac{\partial L}{\partial q}\right|_{q+\Delta q}-\frac{\partial L}{\partial q}
	= 
	\left(\frac{\partial}{\partial q} \frac{\partial L}{\partial q}\right) \Delta q 
	+ \Or\left(|\Delta q|^2\right).
\end{equation}
We thus define an $N \times N$ matrix operator, 
\begin{equation}\label{H(q)}
	H(q) 
	= 
	-\frac{d^2}{dt^2} + \frac{\partial^2}{\partial q^2} L(q,\dot{q}),
\end{equation}
and consider an eigenvalue problem of $H(q)$
\begin{equation}
	H(q)v = \alpha v,
\end{equation}
where $v$ is a column vector of length $N$ and $\alpha$ an eigenvalue. 
Since $H(q)$ is real and symmetric, both $\alpha$ and $v$ are real.

For $\xi \to \xi^{(b)}$, 
by the relation (\ref{hessian}), 
there exist eigenvalues 
\begin{equation}
	\alpha \to 0
\end{equation}
with 
\begin{equation}
	\Delta q \to v.
\end{equation}

Note that, on the other hand, $H(q)$ is related to the second derivative of action as 
\begin{equation}\label{Sv}
	\frac{\partial^2 S(q+xu)}{\partial x^2} = \int_0^T dt u^tH(q)u, 
\end{equation}
for any periodic function $u(t+T)=u(t)$ with $x \in \mathbb{R}$
where 
\begin{equation}\label{def.S}
	S(f) = \int_0^T L(f,\dot{f}) dt
\end{equation}
is the action for a periodic function $f(t+T)=f(t)$ and $^t$ denotes transpose.

\subsection{Representation of bifurcation}
In the vicinity of the bifurcation point $\xi^{(b)}$, 
we represent the bifurcation solution at the fixed $\xi$ as 
\begin{equation}\label{phipsi}
	q^{(b)}=q+\phi r+ \psi \epsilon.
\end{equation} 
Here, 
\begin{equation}\label{phi}
	\phi=(\phi_1, \phi_2, \ldots, \phi_d)
\end{equation}
is an $N \times d$ matrix composed of orthonormal eigenfunctions of $H(q)$
\begin{equation}
	H(q)\phi_i=\kappa_i\phi_i, \; 
	\int_0^T dt \phi_i\phi_j = \delta_{ij},
\end{equation}
for eigenvalues $\kappa_i$ going to zero 
for $\xi \to \xi^{(b)}$, 
and 
\begin{equation}\label{psi}
	\psi=(\psi_1, \psi_2, \ldots) 
\end{equation}
is the orthonormal eigenfunctions for the other, non zero, eigenvalues $\lambda_i$, 
\begin{equation}
	H(q)\psi_i=\lambda_i\psi_i, \; 
	\int_0^T dt \psi_i\psi_j = \delta_{ij},
\end{equation}
where 
$\delta_{ij}$ is Kronecker delta. 
The coefficients $r$ and $\epsilon$ are the column vectors
\begin{equation}
	r={}^t(r_1,r_2,\ldots,r_d),\;
	\epsilon={}^t(\epsilon_1,\epsilon_2,\ldots),
\end{equation}
with $r_i \in \mathbb{R}$ and $\epsilon_i \in \mathbb{R}$.

In this representation by $r$ and $\epsilon$, the variational principle of the action 
\begin{equation}\label{dr}
	\frac{\partial}{\partial r}S(q+\phi r + \psi\epsilon)=0, 
\end{equation}
\begin{equation}\label{de}
	\frac{\partial}{\partial \epsilon}S(q+\phi r + \psi\epsilon)=0,
\end{equation}
is suitable 
for the Euler-Lagrange equation (\ref{eq.motion}).
A solution of (\ref{dr}) and (\ref{de}) at $r \to 0$ for $\xi\to\xi^{(b)}$ 
corresponds to a bifurcation solution $q^{(b)}$ 
and that at $r=0$ the original solution $q$.  

We introduce seven variation functions 
corresponding to infinitesimally small translation of time, 
translation, and infinitesimally small rotation,
\begin{eqnarray}
	\label{qdot}
	\Phi_{1}=\dot{q},\; 
	\Phi_{2}=(e_x,e_x,e_x),\; 
	\Phi_{3}=(e_y,e_y,e_y),\; 
	\Phi_{4}=(e_z,e_z,e_z), 
	\\
	\label{qx} 
	\Phi_{5}=q\times e_x, \;
	\Phi_{6}=q\times e_y, \;
	\Phi_{7}=q\times e_z, \;
\end{eqnarray}
where 
$e_i=(\delta_{xi},\delta_{yi},\delta_{zi})$, 
$q \times e_i = (\bi{r}_1\times e_i,\bi{r}_2\times e_i,\bi{r}_3\times e_i)$, 
and $\times$ is the outer product. 
These are eigenfunctions of $H(q)$ for zero eigenvalue, 
\begin{equation}
	H(q)\Phi_{i}=0,\; i=1,2,\ldots 7
\end{equation}
and do not contribute to the variational principle (\ref{dr}), 
since for any $r$
\begin{equation}
	\frac{\partial}{\partial r} S(q+\Phi_{i}r)=0.
\end{equation}
Thus we exclude the seven eigenfunctions (\ref{qdot})--(\ref{qx}) from (\ref{phi}). 
Then 
in the simple case 
$\phi$ spans eigenspace of $d$ degenerate eigenvalues 
$\kappa_1=\kappa_2=\cdots=\kappa_d$.

The following methods, 
utilizing irreducible representation, the Lyapunov-Schmidt reduction 
are developed in the area of condensed matter physics, particle physics, etc. 
\cite{Chenciner2,golubitsky,golubitsky2,ikeda,sattinger}.

\subsection{The Lyapunov-Schmidt reduction}
The equation (\ref{de}) determines $\epsilon$ as a smooth function $\epsilon(r)$ of a given $r$ by the implicit function theorem since
\begin{equation}
	\frac{\partial^2 S}{\partial \epsilon^2} = \lambda
\end{equation}
where
\begin{equation}
	(\lambda)_{ij}=\lambda_i \delta_{ij}
\end{equation}
with $\lambda_i \ne 0$.

Consequently, by the $\epsilon(r)$, the bifurcation solution (\ref{phipsi}) is represented 
only by $r$, 
\begin{equation}\label{pLS}
	q^{(b)}(r;q)=q+\phi r+\psi \epsilon(r).
\end{equation}
Thus, its action, 
\begin{equation}\label{SLS}
	S(r)=S(q^{(b)}(r;q)),
\end{equation}
and the variational principle (\ref{dr}) and (\ref{de})
\begin{equation}\label{SLSdr}
	\frac{\partial}{\partial r}S(r)=0,
\end{equation}
are also reduced into a $d$-dimensional space of $r$.

\section{Symmetries of bifurcation solutions}\label{sec:symsol}
We denote a symmetry group of a periodic function $f$ as
\begin{equation}
	G(f)=\{g\in G|gf=f\}
\end{equation}
where $G$ is a group of isometric transformations of coordinate, those of time and permutation of bodies acting on the periodic function. 
We will use 
$\sigma \in G$, cyclic permutation of bodies, defined by
\begin{equation}
	\sigma f = \sigma (\bi{r}_1,\bi{r}_2, \bi{r}_3) = (\bi{r}_2,\bi{r}_3, \bi{r}_1),
\end{equation}
$\tau \in G$, exchange of body 1 and 2, by
\begin{equation}
	\tau f = \tau (\bi{r}_1,\bi{r}_2, \bi{r}_3) 
	= (\bi{r}_2,\bi{r}_1, \bi{r}_3),
\end{equation}
$\mu_j \in G$, inversion of $j=x,y,z$ coordinate, by
\begin{equation}
	\mu_j f = (\mu_j\bi{r}_1,\mu_j\bi{r}_2, \mu_j\bi{r}_3), \; 
	\mu_j\bi{r}_i = (s_{xj}x_i,s_{yj}y_i,s_{zj}z_i),\; s_{cj}=1-2\delta_{cj},
\end{equation}
for $f=(\bi{r}_1,\bi{r}_2,\bi{r}_3)$. 
We assume $G(f)$ is a finite group.

For a periodic solution $q$ of (\ref{eq.motion}), 
if all $g \in G(q)$ commute with matrix $H(q)$,
\begin{equation}
	gH(q) = H(q)g,
\end{equation}
the operation by $g \in G(q)$ 
to eigenfunction $\phi r$ of $H(q)$ is given by
a $d$-dimensional representation $D(g)$,   
a $d \times d$ real orthogonal matrix, of $g$ as
\begin{equation}
	g \phi r = \phi D(g) r,
\end{equation}
since $\phi$ is real and orthonormal. 
Thus, $G(\phi r)$, 
a group of symmetries for $\phi r$, is written without $\phi$
\begin{equation}
	G(r;q)=\{g\in G(q) | D(g)r=r\}.
\end{equation}
Further, by operating $g$ in the (\ref{de}) we obtain
\begin{equation}
	\frac{\partial}{\partial (\Pi\epsilon)} S(q+\phi D(q) r + \psi \Pi\epsilon) = 0,
\end{equation}
where $\Pi$ is the representation of $g$ in $\psi$,   
which leads to the well-known theorem \cite{CL,IJ}.
\begin{theorem}\label{lem:pLS}
	If all $g \in G(q)$ commute with $H(q)$,  
	the bifurcation solution $q^{(b)}(r;q)$
	defined by (\ref{pLS}) satisfies
	\begin{equation}\label{SLSsym}
		g q^{(b)}(r;q)=q^{(b)}(D(g)r;q),
	\end{equation}
	where $D(g)$ is a real orthogonal representation of $g$ in $\phi$, 
	that is, $g\phi r=\phi D(g)r$.
\end{theorem}

\begin{corollary}\label{th:symqb}
	If all $g\in G(q)$ commute with $H(q)$,  
	the bifurcation solution $q^{(b)}(r;q)$ is invariant 
	under $g \in G(r;q)$, $g q^{(b)}(r;q)=q^{(b)}(r;q)$, or
	$G\left(q^{(b)}(r;q)\right)=G(r;q)$.  
\end{corollary}

Therefore, 
the representations of $G(q)$ determine the symmetry of the  
bifurcation solution $q^{(b)}(r;q)$ independent of $H(q)$, 
if all $g\in G(q)$ commute with $H(q)$.

\subsection{Symmetries of the figure-eight choreography}
For the figure-eight choreography $q$, 
the $G(q)$ is generated by 
the four operators, 
choreographic shift ${\cal C}$,
\begin{equation}
	{\cal C} q(t) = \sigma q(t-T/3), \label{eq.c}\\
\end{equation}
space and time inversion ${\cal S}$, 
\begin{equation}
	{\cal S} q(t) = -\tau q(-t), \label{eq.s}\\
\end{equation}
$x$-inversion with time shift ${\cal M}$,
\begin{equation}
	{\cal M} q(t) = \mu_x q(t-T/2), \label{eq.m}\\
\end{equation}
and inversion of $z$ coordinate $\mu=\mu_z$, where 
$\bi{r}_3=0$ at $t=0$ is assumed \cite{Chenciner,Chenciner2,fujiwara2020,fukuda2020}.
Since 
\begin{equation}
	{\cal C}^3={\cal S}^2={\cal M}^2=\mu^2=1,\; 
	{\cal CS}={\cal SC}^{-1}
\end{equation}
and
\begin{equation}
	{\cal CM}={\cal MC},\;
	{\cal C}\mu=\mu{\cal C},\;
	{\cal SM}={\cal MS},\;
	{\cal S}\mu=\mu{\cal S},\;
	{\cal M}\mu=\mu{\cal M},
\end{equation}
the group $G(q)$ is isomorphic to a crystallographic point group $D_{6h}$ or  
a direct product of abstract groups $D_6 \times C_2 = D_3 \times C_2 \times C_2$ 
where $C_n$ and $D_n$ are the cyclic and the dihedral group, respectively \cite{pointgroup}. 
Hereafter, we denote generators of the group by curly parenthesis 
following the Schoenflies notation of isomorphic group, then
\begin{equation}
	G(q)=D_{6h}\{{\cal C},{\cal S},{\cal M},\mu\}.
\end{equation}

It is verified that 
the operators $\mathcal{C}$, $\mathcal{S}$, $\mathcal{M}$ and $\mu$ commute with $H(q)$. 
Therefore, the symmetry group $G(r;q)$ for 
bifurcation solution $q^{(b)}(r;q)$ from the figure-eight choreography $q$
is a real orthogonal representations  
of $D_{6h}\{{\cal C},{\cal S},{\cal M},\mu\}$ by corollary~\ref{th:symqb}.

\subsection{Symmetries of the bifurcation solutions from figure-eight choreography}

\begin{table}
	\caption{\label{D6h}
		Symmetries of bifurcation solution $G(r;q)$ 
		for $D_{6h}\{{\cal CM},{\cal S},\mu\} C_{xy}$. 
	}
	\begin{indented}
		\item[]
		\begin{tabular}{cllc}
			\hline
			$\{D({\cal CM}),D({\cal S}),D(\mu)\}$&$G(r;q)$&&\\
			\hline
			$C_1\{+1,+1,+1\}$ & $D_{6h}\{{\cal CM, S},\mu\}$ &&$C_{xy}$\\
			$C_2\{+1,-1,+1\}$ & $C_{6h} \{{\cal CM}, \mu\}$ &&$C_y$\\
			$C_2\{-1,+1,+1\}$ & $D_{6}\{\mu{\cal C}, {\cal S}\}$  &&$C_i$\\
			$C_2\{-1,-1,+1\}$ & $D_{6}\{\mu{\cal C},{\cal SM}\}$ &&$C_x$ \\
			$C_2\{+1,+1,-1\}$ & $D_{6}\{{\cal CM, S}\}$ &&I \\
			$C_2\{+1,-1,-1\}$ & $D_{6}\{{\cal CM}, \mu{\cal S}\}$ &&$\bi{D}_{xy}$\\
			$C_2\{-1,+1,-1\}$ & $D_6\{\mu{\cal CM},{\cal S}\}$ &&I\\
			$C_2\{-1,-1,-1\}$& $D_6\{\mu{\cal CM}, \mu{\cal S}\}$
			&&$\bi{D}_{\hat{x}\hat{y}}$\\
			$D_3\{C(3),F,+E\}$
			&$\left\{ \begin{array}{l} 
				D_{2h}\{\mathcal{C}^{n-1}\mathcal{S},\mathcal{M},\mu\}\\ 
				D_{2}\{\mathcal{M},\mu\} 
			\end{array} \right.$
			&$\begin{array}{l} 
				\displaystyle (\theta=\theta_{2n+1})\\ 
				\mbox{(otherwise)} 
			\end{array}$
			&$\begin{array}{c} D_{xy} \\ D_y \end{array}$
			\\	
			$D_6\{C(3),F,-E\}$
			&$\left\{\begin{array}{l}
				D_{2}\{\mu\mathcal{C}^{n+1}\mathcal{S},{\cal M}\}
				\\
				D_{2}\{\mathcal{C}^{n-1}\mathcal{S},\mathcal{M}\}
				\\ 
				C_{2}\{{\cal M}\} \\
			\end{array} \right.$
			&$\begin{array}{l}
				(\theta=\theta_{2n})\\
				(\theta=\theta_{2n+1})\\
				(\mbox{otherwise})
			\end{array}$
			&$\begin{array}{c}\rm{I}\\ \rm{I}\\ \rm{I} \end{array}$\\
			$D_6\{C(6),F,+E\}$
			&$\left\{ \begin{array}{l}
				D_{2}\{\mathcal{C}^{-n}\mathcal{SM},\mu\}\\
				D_{2}\{\mathcal{C}^{1-n}\mathcal{S}, \mu\}\\
				C_{2}\{\mu\} 
			\end{array} \right.$
			&$\begin{array}{l}
				(\theta=\theta_{2n})\\
				(\theta=\theta_{2n+1})\\
				(\mbox{otherwise})
			\end{array}$
			&$\begin{array}{c} D_{x}\\ D_i\\ D \end{array}$\\
			$D_{6}\{C(6),F,-E\}$&$
			\left\{ \begin{array}{l}
				D_2\{\mathcal{C}^{-n}\mathcal{SM},\mu\mathcal{M}\}\\
				D_2\{\mathcal{C}^{1-n}\mathcal{S},\mu\mathcal{M}\}\\ 
				C_2\{\mu\mathcal{M}\} 
			\end{array} \right.$
			&$\begin{array}{l}
				(\theta=\theta_{2n})\\
				(\theta=\theta_{2n+1})\\
				(\mbox{otherwise})
			\end{array}$
			&$\begin{array}{c} \rm{I}\\ \rm{I}\\ \rm{I} \end{array}$
			\\ 
			\hline
		\end{tabular}	
	\end{indented}
\end{table}
In table \ref{D6h}, 
symmetries of bifurcation solutions, $G(r;q)$, 
for the irreducible real orthogonal representation of 
$D_{6h}\{{\cal C},{\cal S},{\cal M},\mu\}$
are tabulated, where
$D_{nh}=D_n \times C_2$, 
$C_{nh}=C_n \times C_2$, 
$\theta=\arctan(r_2,r_1)$,
\begin{eqnarray}
	\theta_n=\frac{\pi}{6}n,\;
	C(n)=R\left(\frac{2\pi}{n}\right),
	E=	\left(\begin{array}{cc}1&0\\0&1\end{array}\right), 
	F=	\left(\begin{array}{cc}-1&0\\0&1\end{array}\right),
\end{eqnarray}
and 
\begin{equation}
	R(\theta)=\left(\begin{array}{cc}
		\cos\theta&-\sin\theta\\
		\sin\theta&\cos\theta
	\end{array}\right).
\end{equation}
Note that since 
$({\cal CM})^4={\cal C}$ and $({\cal CM})^3={\cal M}$,
${\cal CM}$ is also a generator of the group instead of ${\cal C}$ and ${\cal M}$, 
\begin{equation}
D_{6h}\{{\cal C}, {\cal S}, {\cal M}, \mu\}=D_{6h}\{{\cal CM}, {\cal S}, \mu\}.
\end{equation}

The first column is the irreducible representations of generators of $G(q)$, and  
the second column is the group $G(r;q)$.
For one-dimensional representations the group $G(r;q)$ is given by simply 
$\{g \in  G(q) | D(g)=1\}$, 
and for two-dimensional representations it is derived by the following relation:
For two-dimensional real orthogonal matrix $D$, 
the solution of $Dr=r$ for real $r \ne 0$ is 
\begin{equation}
	r = \left\{ \begin{array}{ll} 
		\mbox{no solution} & (\theta\ne 0, D=R(\theta))\\
		\mbox{all } r & (\theta=0, D=R(\theta))\\
		{}^t(\sin\theta,1+\cos\theta) &(\theta\ne\pi, D=FR(\theta))\\
		{}^t(1,0) &(\theta=\pi, D=FR(\theta)).
	\end{array}\right.
\end{equation}

\subsection{Spatial symmetry of orbit}\label{sec:impossible}
If a periodic motion is symmetric in the reflection across a plane, 
the motion is planar in the plane and vice versa.
In our case since the origin of such symmetry is only $\mu=\mu_z$, 
we could state that if a periodic motion does not have $\mu$ symmetry the motion is not planar.

On the other hand, 
motion with ${\cal S}{\cal C}^x$, $x \in \mathbb{Z}$ symmetry is planar. 
This is because motion with $J=0$ 
is planar \cite{winter,siegel}, and 
${\cal S}J=-J$ and 
${\cal C}J=J$, 
where $J=\sum_j \bi{r}_j \times \dot{\bi{r}}_j$ is angular momentum.  
Therefore, 
the symmetry with ${\cal S}{\cal C}^x={\cal C}^{-x}{\cal S}$ but without $\mu$ is impossible.

In table \ref{D6h}, the spatial symmetry of the shape of the orbit for $q^{(b)}$ 
is tabulated in the last column by the same notation in our previous paper \cite{fukuda2019}. 
The letter $C$ or $D$ distinguishes choreographic or not. 
The subscripts $x$ and $y$ show a symmetric plane where $x$ means $x$-$z$ plane and $y$ $y$-$z$. 
To avoid collision of notation inversion is denoted by subscript $i$ 
instead of $2$ in \cite{fukuda2019}.  
Then we introduced new notations.  
The subscript with a hat shows a symmetric axis of half turn,
the bold letter 
means that the motion is non-planar, and `I' the impossible symmetry discussed above. 
Hereafter we add spatial symmetry 
following full symmetry as $D_{6h}\{{\cal CM},{\cal S},\mu\} C_{xy}$.

\subsection{Symmetry of the sequence of bifurcation solutions}
\begin{table}
	\caption{\label{D6}
		$G(r;q^{(b)})$ 
		for $D_{6}\{m{\cal C},S\}$, \\  
		$(m,S)=(\mu,{\cal S})^{\rm a)}$, 
		$(\mu,{\cal SM})^{\rm b)}$, 
		$({\cal M},\mu{\cal S})^{\rm c)}$, 
		$(\mu{\cal M},\mu{\cal S})^{\rm d)}$. 
	}
	\begin{indented}
		\item[] \begin{tabular}{cllcccc}
		\hline
		\scriptsize $\{D(m{\cal C}),D(S)\}$ & $G(r;q^{(b)})$ & 
		&\scriptsize a)&\scriptsize b)&\scriptsize c)&\scriptsize d)
		\\ \hline
		$C_1\{+1,+1\}$ & $D_{6}\{m{\cal C}, S\}$&
		& $C_i$& $C_y$& $\bi{C}_{xy}$& 
		$\bi{C}_{\hat{x}\hat{y}}$\\  
		$C_2\{+1,-1\}$ & $C_{6} \{m{\cal C}\}$  &
		&$C$ & $C$& $\bi{C}_{y}$
		& $\bi{C}_{\hat{y}}$\\
		$C_2\{-1,+1\}$ & $D_{3}\{{\cal C}, S\}$ &
		& I
		& $\bi{C}_{\hat{x}}$& $\bi{C}_{\hat{z}}$& $\bi{C}_{\hat{z}}$\\
		$C_2\{-1,-1\}$ & $D_{3}\{{\cal C},mS\}$	&
		& $\bi{C}_{\hat{z}}$& $\bi{C}_{\hat{y}}$ & $\bi{C}_{x}$& $\bi{C}_{\hat{x}}$\\
		$D_3\{C(3),F\}$&
		$\left\{ \begin{array}{l} 
			D_{2}\{{\cal C}^{n-1}S,m\}\\ 
			C_{2}\{m\} 
		\end{array} \right.$	
		&$\begin{array}{l} 
			(\theta=\theta_{2n+1})\\ 
			(\mbox{otherwise}) 
		\end{array}$
		&$\begin{array}{c} D_i \\ D \end{array}$
		&$\begin{array}{c} D_y \\ D \end{array}$
		&$\begin{array}{c} \bi{D}_{xy} \\ \bi{D}_{y} \end{array}$
		&$\begin{array}{c} \bi{D}_{\hat{x}\hat{y}} \\ \bi{D}_{\hat{y}} \end{array}$	
		\\
		$D_6\{C(6),F\}$
		&$\left\{ \begin{array}{l}
			C_{2}\{m{\cal C}^{-n}S\}\\
			C_{2}\{{\cal C}^{1-n}S\}\\ 
			C_1\{1\} 
		\end{array} \right.$
		&$\begin{array}{l}
			(\theta=\theta_{2n})\\
			(\theta=\theta_{2n+1})\\
			(\mbox{otherwise})
		\end{array}$
		&$\begin{array}{c} \rm{I}\\ \rm{I}\\ \rm{I} \end{array}$
		&$\begin{array}{c} \bi{D}_{x}\\ \bi{D}_i\\ \bi{D} \end{array}$
		&$\begin{array}{c} \bi{D}_{x}\\ \bi{D}_{\hat{z}}\\ \bi{D} \end{array}$
		&$\begin{array}{c} \bi{D}_{\hat{x}}\\ \bi{D}_{\hat{z}}\\ \bi{D}\end{array}$
		\\ \hline
		\end{tabular}
	\end{indented}
\end{table}
In table \ref{D6}, the symmetry $G(r;q^{(b)})$ of 
the bifurcation solution from $q^{(b)}$ with $G(q^{(b)})=D_6$, 
the second bifurcation solutions from 
the figure-eight choreography $q$ through $D_6$, 
are shown in the same manner.
There are six $D_6$ representations in table \ref{D6h}, 
but two of them are impossible indicated by `I', and the other 
four,
$D_6\{\mu{\cal C},{\cal S}\}C_i$,   
$D_6\{\mu{\cal C},{\cal SM}\}C_x$,   
$D_6\{{\cal CM},\mu{\cal S}\}\bi{D}_{xy}$ 
and 
$D_6\{\mu{\cal CM},\mu{\cal S}\}\bi{D}_{\hat{x}\hat{y}}$   
are listed in the caption 
and represented as a)--d)
in table \ref{D6}.  
The spatial symmetry of $G(r;q^{(b)})$ of these groups are tabulated in the last four columns titled by a)--d).

\begin{table}
	\caption{\label{D2h} $G(r;q^{(b)})$ for}
	\begin{indented}
	\item[]	
	\begin{tabular}{clc}
	\multicolumn{3}{l}{(a) $C_{6h}\{{\cal CM},\mu\}C_y$,} 
	\\ \hline
	$\{D({\cal CM}),D(\mu)\}$ & $G(r;q^{(b)})$&\\
	\hline
	$C_1\{+1,+1\}$ & $C_{6h}\{{\cal CM}, \mu\}$& $C_y$\\
	$C_2\{-1,+1\}$ & $C_{6}\{\mu{\cal C}\}$& $C$\\
	$C_2\{+1,-1\}$ & $C_{6} \{{\cal CM}\}$& $\bi{C}_{y}$\\
	$C_2\{-1,-1\}$ & $C_{6}\{\mu{\cal CM}\}$& $\bi{C}_{\hat{y}}$\\
	$C_3\{C(3),+E\}$ & $D_{2}\{{\cal M},\mu\}$& $D_y$\\
	$C_6\{C(3),-E\}$ & $C_{2}\{{\cal M}\}$& $\bi{D}_{y}$\\
	$C_6\{C(6),+E\}$ & $C_{2}\{\mu\}$& $D$\\
	$C_{6}\{C(6),-E\}$ & $C_{2}\{\mu{\cal M}\}$& $\bi{D}_{\hat{y}}$\\
	\hline
\end{tabular}	
	\begin{tabular}{clc}
	\multicolumn{3}{l}{(b) $D_{2h}\{S,{\cal M},\mu\}D_{xy},S={\cal C}^{n-1}{\cal S}$.} 
	\\ \hline
	\scriptsize$\{D(S),D({\cal M}),D(\mu)\}$ & $G(r;q^{(b)})$&\\
	\hline
	$C_1\{+1,+1,+1\}$ & $D_{2h}\{S, {\cal M},\mu\}$& $D_{xy}$\\  
	$C_2\{+1,-1,+1\}$ & $D_{2}\{S, \mu\}$& $D_{i}$\\
	$C_2\{-1,+1,+1\}$ & $D_{2}\{{\cal M}, \mu\}$& $D_{y}$\\
	$C_2\{-1,-1,+1\}$ & $D_{2}\{S{\cal M},\mu\}$& $D_{x}$\\
	$C_2\{+1,+1,-1\}$ & $D_{2}\{S, {\cal M}\}$& I\\  
	$C_2\{+1,-1,-1\}$ & $D_{2} \{S, \mu{\cal M}\}$& I\\
	$C_2\{-1,+1,-1\}$ & $D_{2}\{\mu S,{\cal M}\}$& $\bi{D}_{xy}$\\
	$C_2\{-1,-1,-1\}$ & $D_{2}\{\mu S,\mu{\cal M}\}$& $\bi{D}_{\hat{x}\hat{y}}$\\
	\hline
	\end{tabular}
	\end{indented}
\end{table}
\begin{table}
	\caption{\label{D2}
		$G(r;q^{(b)^k})$, $k=1,2$ for 
		$D_{2}\{S,m\}$, 
		$(S,m)=$ 
		$({\cal C}^{n}{\cal S},\mu)^{\rm a)}$,
		$({\cal C}^{n}{\cal SM},\mu)^{\rm b)}$,
		$({\cal M},\mu)^{\rm c)}$, 
		$(\mu{\cal C}^{n}{\cal S},{\cal M})^{\rm d)}$,
		$(\mu{\cal C}^{n}{\cal S},\mu{\cal M})^{\rm e)}$.
	}
	\begin{indented}
	\item[]	
	\begin{tabular}{clccccc}
		\hline
		$\{D(S),D(m)\}$ & $G(r;q^{(b)^k})$ 
		&\scriptsize a)&\scriptsize b)&\scriptsize c)&\scriptsize d)&\scriptsize e) \\
		\hline
		$C_1\{+1,+1\}$ & $D_{2}\{S,m\}$& $D_i$& $D_x$& $D_y$& $\bi{D}_{xy}$& 
		$\bi{D}_{\hat{x}\hat{y}}$\\  
		$C_2\{+1,-1\}$ & $C_{2}\{S\}$ & I
		& $\bi{D}_{\hat{x}}$& $\bi{D}_{y}$& $\bi{D}_{\hat{z}}$& $\bi{D}_{\hat{z}}$ \\
		$C_2\{-1,+1\}$ & $C_{2}\{m\}$ 
		& $D$& $D$& $D$& $\bi{D}_y$& $\bi{D}_{\hat{y}}$\\
		$C_2\{-1,-1\}$ & $C_{2}\{m S\}$ & $\bi{D}_{\hat{z}}$& $\bi{D}_{x}$& $\bi{D}_{\hat{y}}$& $\bi{D}_{x}$& $\bi{D}_{\hat{x}}$ \\
		\hline
	\end{tabular}	
	\end{indented}
\end{table}
In tables \ref{D2h} and \ref{D2}, 
symmetries of the bifurcation solutions 
from $q$ through $C_{6h}$, $D_{2h}$ and $D_{2}$, that is  
the other second bifurcation solutions, are shown. 
Though table \ref{D2h} contains only direct bifurcations 
through $C_{6h}$ and $D_{2h}$ in table \ref{D6h}, table \ref{D2} contains 
not only $D_2$ in table \ref{D6h} 
but also $D_2$ in tables \ref{D6} and \ref{D2h}, 
that is the third bifurcation $q^{(b)^2}=(q^{(b)})^{(b)}$ from $q$.
 
\begin{table}
	\centering
	\caption{\label{D3}
		$G(r;q^{(b)^2})$ for 
		$D_{3}\{{\cal C},S\}$, 
		$S=\mu{\cal S}^{\rm a)}, {\cal S}{\cal M}^{\rm b)}, \mu{\cal S}{\cal M}^{\rm c)}$.
	}
	\begin{indented}
	\item[] 
	\begin{tabular}{cllccc}
		\hline
		$\{D(\mu{\cal C}),D(S)\}$ & $G(r;q^{(b)^2})$ & 
		&\scriptsize a)&\scriptsize b)&\scriptsize c)\\
		\hline
		$C_1\{+1,+1\}$ & $D_{3}\{{\cal C}, S\}$&
		& $\bi{C}_{\hat{z}}$& $\bi{C}_{\hat{x}}$& $\bi{C}_{x}$\\  
		$C_2\{+1,-1\}$ & $C_{3} \{{\cal C}\}$&
		& $\bi{C}$& $\bi{C}$& $\bi{C}$\\
		$D_3\{C(3),F\}$&
		$	\left\{ \begin{array}{l} 
			C_{2}\{\mathcal{C}^{n-1}S\}\\ 
			C_{1}\{1\} 
		\end{array} \right.$	
		&
		$	\begin{array}{l} 
			\displaystyle (\theta=\theta_{2n+1})\\ 
			(\mbox{otherwise}) 
		\end{array}
		$&$\begin{array}{c} \bi{D}_z\\ \bi{D} \end{array}$
		&$\begin{array}{c} \bi{D}_{\hat{x}} \\ \bi{D} \end{array}$
		&$\begin{array}{c} \bi{D}_{x}\\ \bi{D} \end{array}$			
		\\ \hline
	\end{tabular}
	\end{indented}
\end{table}
\begin{table}
	\caption{\label{C6}
		$G(r;q^{(b)^2})$ for  
		$C_{6}\{m{\cal C}\}$, 
		$m=\mu^{\rm a)},{\cal M}^{\rm b)}, \mu{\cal M}^{\rm c)}$.
	}
	\begin{indented}
		\item[] 
		\begin{tabular}{clccc}
			\hline
			$\{D(m{\cal C})\}$ & $G(r;q^{(b)^2})$ &\scriptsize a)&\scriptsize b)&\scriptsize c)\\
			\hline
			$C_1\{+1\}$ & $C_{6}\{m{\cal C}\}$ 
			&$C$&$\bi{C}_y$&$\bi{C}_{\hat{y}}$\\  
			$C_2\{-1\}$ & $C_{3}\{{\cal C}\}$ &$\bi{C}$&$\bi{C}$&$\bi{C}$\\
			$C_3\{C(3)\}$ & $C_{2}\{m\}$&$D$&$\bi{D}_y$&$\bi{D}_{\hat{y}}$\\
			$C_6\{C(6)\}$ & $C_{1}\{1\}$&$\bi{D}$&$\bi{D}$&$\bi{D}$\\
			\hline
		\end{tabular}
	\end{indented}
\end{table}
Further, in tables \ref{D3} and \ref{C6}, 
symmetries $G(r;b^{(b)^2})$ of the third bifurcation solutions from $q$ 
through $D_6 \to D_3$, $D_6 \to C_6$ are shown. 
\begin{table}
	\caption{\label{C2}	$G(r;q^{(b)^k})$, $k=2,3$ for}
	\begin{indented}
		\item[] 
		\begin{tabular}{clc}
			\multicolumn{3}{l}{(a) $C_{3}\{{\cal C}\}C$,} 
			\\
			\cline{1-3}
			$\{D({\cal C})\}$ & $G(r;q^{(b)^3})$&\\
			\cline{1-3}
			$C_1\{+1\}$ & $C_{3}\{{\cal C}\}$& $\bi{C}$ \\  
			$C_3\{C(3)\}$ & $C_{1}\{1\}$& $\bi{D}$ \\
			\cline{1-3}
		\end{tabular}
		\begin{tabular}{clcccccccc}
			\multicolumn{10}{l}{(b) $C_{2}\{m\}$, 
				$m=\mu^{\rm a)}, {\cal M}^{\rm b)}, \mu{\cal M}^{\rm c)}, \mu{\cal SM}^{\rm d)}, \mu{\cal S}^{\rm e)}, {\cal SM}^{\rm f)}$.}\\
			\cline{1-8}
			$\{D(m)\}$ & $G(r;q^{(b)^k})$&\scriptsize a)&\scriptsize b)&\scriptsize c)&\scriptsize d)&\scriptsize e)&\scriptsize f)\\
			\cline{1-8}
			$C_1\{+1\}$ & $C_{2}\{m\}$
			& $D$& $\bi{D}_y$& $\bi{D}_{\hat{y}}$& $\bi{D}_x$& $\bi{D}_i$& $\bi{D}_{\hat{x}}$\\  
			$C_2\{-1\}$ & $C_{1}\{1\}$
			& $\bi{D}$& $\bi{D}$& $\bi{D}$& $\bi{D}$& $\bi{D}$& $\bi{D}$\\
			\cline{1-8}
		\end{tabular}
	\end{indented}
\end{table}
Finally in table \ref{C2},  
the fourth $q^{(b)^3}=(q^{(b)^2})^{(b)}$ through $C_3$, 
and the third and the fourth through $C_2$ are shown.

These sequences of bifurcations stop at 
the bifurcation solution with no symmetry, $C_1\{1\}D$.
Tables \ref{D6h}--\ref{C2}, thus, cover the all sequence of bifurcations from 
the figure-eight choreography $q$ through the irreducible representations. 
\begin{figure}
	\centering
	\includegraphics[width=16cm]{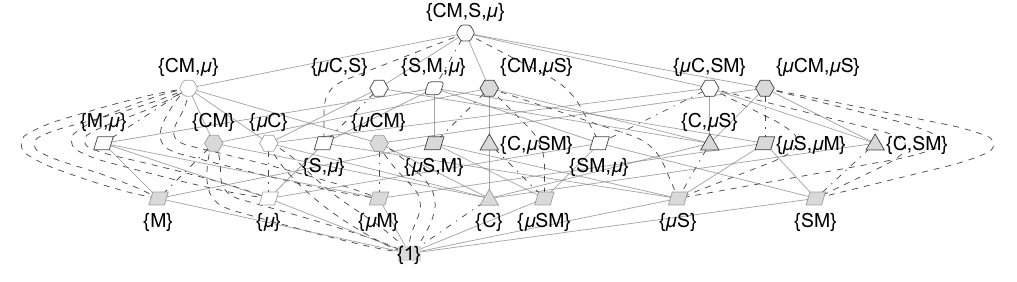}
	\\
	\caption{\label{fig:gr}
		Diagram for cascade of bifurcations. 
		Shapes with $n$-fold symmetry show groups $D_{n(h)}$ or $C_{n(h)}$. 
		The symbol with the dark edge is $D_{n(h)}$, the light edge $C_{n(h)}$,  
		the rounded edge $n$ accompanying $h$, and  
		the grayed face non-planar. 
		$n \ge 3$ is choreographic.
		Lines connecting symbols show fold $n$ of representation 
		$D_{n}$ or $C_{n}$. 
		The solid line is $n=2$, dot-dashed 3, and dashed 6, respectively. 
		Factor ${\cal C}^x$ in generators is omitted.
	}
\end{figure}

Figure \ref{fig:gr} shows a diagram of the bifurcations, 
that is, a diagram of tables \ref{D6h}--\ref{C2}. 
Shapes with $n$-fold symmetry show the bifurcation solutions 
with symmetries $D_{n(h)}$ or $C_{n(h)}$. 
Here and hereafter $D_{n(h)}$ means $D_{n}$ or $D_{nh}$, and $C_{n(h)}$ 
$C_{n}$ or $C_{nh}$.
The shape with the dark edge is $D_{n(h)}$, the light edge $C_{n(h)}$,  
the rounded edge $n$ accompanying $h$, and the grayed face is non-planar. 
Note that solutions with $n \ge 3$ are choreographic and vice versa.

Lines connecting symbols show fold $n$ of the group 
$D_{n}$ or $C_{n}$ 
isomorphic to the representation shown in the first column in tables \ref{D6h}--\ref{C2}, 
corresponding to bifurcation type defined in section \ref{sec:symact}.
The solid line is $n=2$, dot-dashed 3, and dashed 6, respectively.

\section{Symmetries of the action}\label{sec:symact}
If all $g \in G(q)$ commute with $H(q)$ and a $g \in G(q)$ is an invariance of the action, 
that is 
\begin{equation}\label{Sgf}
	S(gf)=S(f)
\end{equation}
for any periodic function $f$, the action for the main term of the   
bifurcation solution has symmetry
\begin{equation}
	S(q+\phi r)=S(g(q+\phi r))=S(q+\phi D(g)r).
\end{equation}
Further by theorem~\ref{lem:pLS},  
\begin{eqnarray}\label{SLSr}
	S(r)&=S\left(q^{(b)}(r;q)\right)
	=S\left(g q^{(b)}(r;q)\right)
	=S\left( q^{(b)}(D(g)r;q) \right)\nonumber \\
	&=S\left( D(g)r \right).
\end{eqnarray}
\begin{corollary}\label{th:SLS}
If all $g \in G(q)$ commute with $H(q)$	and a $g \in G(q)$ is an 
invariance of action (\ref{Sgf}),
	the action $S(r)$ defined by (\ref{SLS}) has a symmetry
	\begin{equation}\label{SLSsym}
		S(r)=S\left(D(g)r\right).
	\end{equation}
\end{corollary}

Since all $g= {\cal C},{\cal S},{\cal M}$ and $\mu$ 
are invariances of the action (\ref{Sgf}), 
corollary~\ref{th:SLS} determines the structure of $S(r)$ and of bifurcation
as in the following subsections \cite{CL}. 
We denote a group of representations for $G(q)$ as
\begin{equation}\label{D(q)}
	D(q)=\{D(g)|g\in G(q)\}, 
\end{equation} 
and inversely set of elements corresponding to $D \in D(q)$ as 
\begin{equation}\label{g(D)}
	g(D;q)=\{g \in G(q) | D(g)=D\}.
\end{equation}

\subsection{One-dimensional bifurcation}\label{sec:1d}
We consider the case $d=1$.
The bifurcation corresponding to the non-degenerate eigenvalue 
of $H(q)$ is this type.
We denote $r_1$ by $r$ and $\kappa_1$ by $\kappa$.

\subsubsection{One-fold type (Trivial type)} 
If $D(q)=C_1\{1\}$, 
that is, trivial representation,
\begin{equation}\label{SLSk1}
	S(r)=S(q)+\frac{\kappa}{2} r^2 + \frac{A_3}{3!} r^3 + \Or(r^4)
\end{equation}
where $A_3$ is a constant independent of $r$.
Suppose $A_3\ne 0$, solution for (\ref{SLSdr})
at $r \ne 0$ with $r\to 0$ for $\kappa\to 0$ is given by
\begin{equation}
	r = -\frac{2}{A_3}\kappa +\Or(\kappa^{2}).
\end{equation}
Thus, there exist two bifurcation solutions with the same symmetry $G(r;q)=G(q)$
in the both sides of $\kappa$,
$\kappa>0$ and $\kappa<0$, with the action value
\begin{equation}
	S(\kappa)=S(q)+\frac{2}{3A_3^2}\kappa^3 +\Or(\kappa^4).
\end{equation} 

In this case, sometimes the relation between $\xi$ and $\kappa$ is quadratic,   
$\xi = \xi^{(b)} + a \kappa^2 +\Or(\kappa^3)$, 
which gives fold bifurcation in the one side of
parameter $\xi$, $\xi>\xi^{(b)}$ or $\xi<\xi^{(b)}$.

\subsubsection{Two-fold type}
If $D(q)=C_2\{1,-1\}$, $S(-r)=S(r)$ and
\begin{equation}\label{S2}
	S(r)=S(q)+\frac{\kappa}{2} r^2+\frac{A_4}{4!}r^4+\Or(r^6)
\end{equation}
where $A_4$ is a constant independent of $r$.
Suppose $A_4\ne 0$, solution for (\ref{SLSdr})
at $r \ne 0$ with $r\to 0$ for $\kappa\to 0$
is given by
\begin{equation}
	r=\pm\sqrt{\frac{-6\kappa}{A_4}}+\Or(\kappa).
\end{equation}
Thus, there exist two bifurcation solutions with the same symmetry $G(r;q)$, 
and with the same action value 
\begin{equation}
	S(\kappa)=S(q)-\frac{3}{2A_4}\kappa^2 +\Or(\kappa^3)
\end{equation}
in the $\kappa/A_4<0$ side of 
the bifurcation point.
If we denote one of the two bifurcation solutions by $q^{(b)}$, 
the other is given by $g_c q^{(b)}$, $g_c \in g(-1;q)$ 
by theorem~\ref{lem:pLS}. 
Thus, two solutions have a congruent orbit.

\subsection{Two-dimensional bifurcation}\label{sec:2d}

We consider the case $d=2$.
We assume that $\kappa_1=\kappa_2=\kappa$ and the representation is irreducible.
Bifurcation corresponding to a 
doubly degenerate eigenvalue of $H(q)$ will be of this type,
at least in all our numerical calculations \cite{fukuda2019,fukuda2019p,fukuda2023}.
We denote $r=|(r_1,r_2)|$ and $\theta=\arctan(r_2,r_1)$.

\subsubsection{Three-fold type}\label{sec:3fold}
If $D(q)$ is isomorphic to $D_3$ or $C_3$, 
\begin{eqnarray}\label{S3}
	S(r,\theta)
	=& S(q) +\frac{\kappa}{2}r^2 +\frac{A_3'\sin (3\theta+3\varphi_3')}{3!}r^3 
	+\frac{A_4^{(0)}}{4!}r^4 
	+ \Or(r^5)
\end{eqnarray}
where 
$A_3'$, $A_4^{(0)}$ and $\varphi_3'$ are constants independent of both $r$ and $\theta$, and $\varphi_3'=0$ if $D(q)=D_3$.
Suppose $A_3' \ne 0$, $r > 0$ solution for (\ref{SLSdr})
with $r\to 0$ for $\kappa\to 0$ is given by
\begin{eqnarray}
	r = \left|\frac{2\kappa}{A_3'}\right|+\Or(\kappa^2), \; 
	\theta = \left\{\begin{array}{ll}
		\displaystyle \theta_{4n+1}-\varphi_3 &(\kappa>0)\\
		\displaystyle \theta_{4n-1}-\varphi_3 &(\kappa<0)\\
	\end{array}\right.,\;
	\varphi_3=\varphi_3'+\Or(\kappa^2)
	\label{theta3} 
\end{eqnarray}
where 
$\Or(\kappa^2)$ in $\varphi_3$ does not depend on $n$. 
Further
if $D(q)=D_3$, $\Or(\kappa^2)=\varphi_3=0$. 
Thus, there exist six bifurcation solutions with the action value
\begin{equation}
	S(\kappa)=S(q)+\frac{2}{3(A_3')^2}\kappa^3+\Or(\kappa^4).
\end{equation} 
The three solutions in the $\kappa>0$ side 
have the symmetry $G(r,\theta_{4n+1}-\varphi_3;q)$ 
and three solutions in the $\kappa<0$ side 
$G(r,\theta_{4n-1}-\varphi_3;q)$.
On one side, 
if we denote one of the three bifurcation solutions by $q^{(b)}$, 
the rest two are given by $g_c^jq^{(b)}$, $j=1,2$, 
$g_c \in g(C(3);q)$, by theorem~\ref{lem:pLS}. 
Thus, the three solutions have a congruent orbit.

\subsubsection{Six-fold type}\label{sec:6fold}
If $D(q)$ is isomorphic to $C_{6}$ or $D_{6}$, 
\begin{eqnarray}\label{S6}
	S(r,\theta)=&S(q)+\frac{\kappa}{2}r^2
	+\frac{A_4^{(0)}}{4!}r^4 
	+\frac{A_6' \cos(6\theta+6\varphi_6')+A_6''}{6!}r^6
	+\Or(r^{8})
\end{eqnarray}
where
$A_4^{(0)}$, $A_6'$, $A_6''$ and $\varphi_6'$ are constants independent of both $r$ and $\theta$, and $\varphi'_6=0$ if 
$D(q)=D_{6}$.
Suppose $A^{(0)}_4 A_6'\ne 0$, $r > 0$ solution for (\ref{SLSdr})
with $r\to 0$ for $\kappa\to 0$ is given by
\begin{equation}
	r = \sqrt{-\frac{6\kappa}{A^{(0)}_4}}+\Or(\kappa^{\frac{3}{2}}), \;
	\theta = \displaystyle \theta_n -\varphi_6, \; 
	\varphi_6=\varphi_6'+ \Or(\kappa),  
	\label{theta6}
\end{equation}
where 
$\Or(\kappa)$ in $\varphi_6$ does not depend on $n$. 
Further if 
$D(q)=D_{6}$, 
$\Or(\kappa)=\varphi_6=0$.
Thus, there exist twelve bifurcation solutions, in the $\kappa/A^{(0)}_4<0$ side. 

The six solutions at $\theta=\theta_{2n}-\varphi_6$  
with the symmetry $G(r,\theta_{2n}-\varphi_6;q)$ have the same action value,
and the other six at $\theta=\theta_{2n+1}-\varphi_6$  
with the symmetry $G(r,\theta_{2n+1}-\varphi_6;q)$ 
have the same action value.
Their action values are slightly different as 
\begin{equation}
	S(\kappa)=S(q)-\frac{3}{2A_4^{(0)}}\kappa^2
	-\frac{3((-1)^nA_6'+A_6'')}{10(A_4^{(0)})^3}\kappa^3+\Or(\kappa^4).
\end{equation}
If we denote one of the six bifurcation solutions with the same action value 
by $q^{(b)}$, the rest five are $g_c^jq^{(b)}$, 
$g_c \in g(C(6);q)$, $j=1,2,\ldots,5$ by theorem~\ref{lem:pLS}. 
Thus, the six solutions have a congruent orbit. 
Note that $(-C(3))^5=C(6)$.

In tables \ref{D6h} and \ref{D6}, we classified a 
six-fold representation of $D_6$ as 
impossible if symmetry at either $\theta_{2n}$ or $\theta_{2n+1}$ is impossible. 
This is because both $\theta_{2n}$ and $\theta_{2n+1}$ are critical 
points, as shown in (\ref{theta6}).

\section{Numerical solutions}\label{sec:num}
Numerical solutions were calculated by Mathematica 14 \cite{mathematica}. 
Periodic solutions with symmetries are searched according to the method described in \cite{fukuda2019,fukuda2020}.
The solutions without symmetry are 
searched as follows: 
Solve the Euler-Lagrange equation (\ref{eq.motion}) 
in the center of mass system with initial conditions by 
seven (five in planar motion) parameters $\bi{r}_2(0)$, $|\dot{\bi{r}}_1(0)|$, $\dot{\bi{r}}_2(0)$ 
for fixed $\bi{r}_1(0)$ and direction of $\dot{\bi{r}}_1(0)$.
Then demand the same number of conditions 
$\bi{r}_1(0)=\bi{r}_1(T)$, 
$|\dot{\bi{r}}_1(0)|=|\dot{\bi{r}}_1(T)|$, 
$\bi{r}_2(0)=\bi{r}_2(T)$ 
by the Newton method. 

In the vicinity of the bifurcation point, 
there are at least two incongruent solutions, including the original solution. 
The method equivalent to that described in the reference \cite{doedel,doedel2} 
is used to resolve several close solutions.
Moreover, the method is coded in a set of FORTRAN subroutines named AUTO \cite{auto}. 
However, we did not use it since we started our calculations with Mathematica.
We believe that our numerical calculation can be done effectively by AUTO.

In the calculation of the Euler-Lagrange equation (\ref{eq.motion}) and 
eigenvalue problem of $H(q)$, 
instead of $q=(\bi{r}_1,\bi{r}_2,\bi{r}_3)$ we use the Jacobi coordinate,
\begin{equation}
	Q=(R_1,R_2),\; R_1=\bi{r}_2-\bi{r}_1, R_2=\bi{r}_3-\frac{\bi{r}_1+\bi{r}_2}{2},
\end{equation}
which separates the center of mass and reduces 
the number of variables from $9$ to $6$ 
($6$ to $4$ for the planar problem). 
Then results are displayed by inverse relation
\begin{equation}
	q=(-\frac{1}{2}R_1-\frac{1}{3}R_2,
	\frac{1}{2}R_1-\frac{1}{3}R_2, 
	\frac{2}{3}R_2).
\end{equation}

In order to find bifurcation, first, we calculate the eigenvalues of $H(q^{(b)^k})$ 
by truncated Fourier expansions \cite{fukuda2018} and search for their zero. 
Then in the vicinity of zero, identify the symmetry $G(q^{(b)^k}+\phi r)$ 
of a function $q^{(b)^k}+\phi r$
composed of their eigenfunctions by the plot of $q^{(b)^k}+\phi r$ 
with suitably large $r$ to emphasize spatial symmetry \cite{fukuda2019}.
Since the last several columns on tables \ref{D6h}--\ref{C2} showing spatial symmetry of $q^{(b)^{k+1}}$ do not contain the same spatial symmetry in the same column, 
we can identify it. 
Then the full symmetry $G(q^{(b)^k}+\phi r)=G(r; q^{(b)^k})$ is determined from the corresponding second column. 
The symmetry of the bifurcation solution 
$G(q^{(b)^{k+1}}(r;q^{(b)^k}))$ is $G(q^{(b)^k}+\phi r)$ 
by corollary~\ref{th:symqb}, and the bifurcation type,  
one of the four types described in section \ref{sec:1d} and \ref{sec:2d}, is identified  
in the first column.
We then search for a bifurcation solution effectively  
by Newton method described above assuming the symmetry of the solution. 

In order to ensure the existence of bifurcation solutions, 
we have to investigate the coefficients 
in (\ref{SLSk1}), (\ref{S2}), (\ref{S3}) and (\ref{S6}).
However, we skip this process and assume the existence of bifurcation solutions, 
then search them numerically by Newton method explained above.
If the numerical bifurcation solutions are found and their symmetries coincide, 
we regard them as the bifurcation solutions expected by the theory, 
though, of course, this is not a rigorous way.

\subsection{The Lennard-Jones potential system}
\begin{table}
	\centering
	\caption{\label{LJ}
		$q^{(b)}$ from $\alpha^{\pm}$ solution $q$ 
		with $D_{6h}\{{\cal CM},{\cal S},\mu\}C_{xy}$ 
		for LJ system. 
	}
	\begin{indented}
		\item[]	\begin{tabular}{cccclc}
		\hline
		$T^{(b)}$&branch& $\kappa$&$D$&$G(q^{(b)})$ 
		& side
		\\ \hline
		$14.861$&$-$&$\swarrow$& $D_6$
		&$D_{2}\{{\cal C}^{-n}{\cal SM},\mu\}D_x > D_{2}\{{\cal C}^{1-n}{\cal S},\mu\}D_i$ 
		& R\\
		$14.836$&$-$&$\swarrow$&$D_3$
		&$D_{2h}\{{\cal C}^{n-1}{\cal S},{\cal M},\mu\}D_{xy}$  
		&B\\
		$14.595$&$-$&$\swarrow$&$C_2$
		&$D_{6}\{\mu{\cal C},{\cal SM}\}C_y$ 
		&R\\
		$14.479$&$0$&$(\uparrow$&$C_1$
		&$D_{6h}\{{\cal CM},{\cal S},\mu\}C_{xy}$ 
		&B\\ 
		$16.111$&$+$&$\searrow$& $C_6$ 
		&$D_{2}\{{\cal C}^{-n}{\cal SM},\mu\}D_x < D_{2}\{{\cal C}^{1-n}{\cal S},\mu\}D_i$ 
		&R\\
		$16.878$&$+$&$\searrow$&$D_3$
		&$D_{2h}\{{\cal C}^{n-1}{\cal S},{\cal M},\mu\}D_{xy}$ 
		&B \\
		$17.132$&$+$&$\searrow$&$C_2$
		&$C_{6h}\{{\cal CM},\mu\}C_y$ 
		&R \\
		$18.615$&$+$&$\searrow$&$C_2$
		&$D_{6}\{\mu{\cal C},{\cal S}\}C_i$ 
		&R \\
		\hline
		\end{tabular}	
	\end{indented}
\end{table}
In table \ref{LJ}, we show all bifurcations calculated numerically in \cite{fukuda2019} 
for the system with the Lennard-Jones (LJ) potential
\begin{equation}
	u(r)=-\frac{1}{r^6}+\frac{1}{r^{12}}.
\end{equation}
The bifurcation parameter $\xi$ is the period $T$ and 
the bifurcation point $\xi^{(b)}$ is $T^{(b)}$ tabulated in the first column. 

The second column, `branch', shows the sign of eigenvalue  
for trivial representation corresponding to fold bifurcation of original solution $q$ itself. 
At the point denoted by 0, trivial type bifurcation occurs,  
and the branch of $q$ changes.
We called the `$-$' branch $\alpha^-$ solution 
and `$+$' $\alpha^+$ \cite{fukuda2017}.  
The $T^{(b)}$ in the first column is 
tabulated in the order tracing branches in turn.  

The third column, `$\kappa$', shows the tangent vector of the  
curve $(T,\kappa(T))$ for the bifurcation solution $q^{(b)}$ with $T$ 
varying in the order of $T^{(b)}$ in the first column.
A single parenthesis accompanied by an up or down vector shows the 
derivative of a tangent.

The groups of irreducible representations and the symmetries of 
bifurcation solutions $q^{(b)}$
are shown in the columns `$D$' and `$G(q^{(b)})$', respectively.
The bifurcation type is given by the subscript $n$ 
of 
$D_{n}$ or $C_{n}$ 
in the column `D' as an $n$-fold type. 
An inequality between symmetries in the column `$G(q^{(b)})$'
is the magnitude relation of their action values. 

The last column, `side', shows the side that bifurcation solution exists. 
`L', `R' and `B' mean left, right, and both sides, respectively.  
Note that `B' is determined by 
the eigenvalue $\kappa$ and eigenfunctions $\phi$ without bifurcation solution, 
whereas, in order to determine `L' or `R'
and also inequality in the column `$G(q^{(b)})$'
numerical solution is necessary.

\begin{table}
	\centering
	\caption{\label{LJ18615}
		$q^{(b)^2}$ from $q^{(b)}$ with $D_{6}\{\mu{\cal C},{\cal S}\}C_i$  
		from $\alpha^+$ solution $q$ in table \ref{LJ}.
	}
	\begin{indented}
		\item[]	\begin{tabular}{cccclc}
			\hline
			$T^{(b)^2}$&branch&$\kappa$&$D$ & $G(q^{(b)^2})$ 
			& side\\ 
			\hline
			$18.615$&$0$&$\nearrow$&$C_1$&$D_{6}\{\mu{\cal C},{\cal S}\} C_i$
			&R	\\ 		 
			$19.020$&$+$&$\nearrow$&$C_2$&$C_{6}\{\mu{\cal C}\}C$ 
			&R \\ 		 
			$19.250$&$+$&$\nearrow$&$D_3$&$D_{2}\{{\cal C}^{n-1}{\cal S},\mu\} D_i$ 
			&B	\\ 		 
			$20.100$&$+$&$\nearrow$&$D_3$&$D_{2}\{{\cal C}^{n-1}{\cal S},\mu\} D_i$ 
			&B	\\ 		 
			\hline
		\end{tabular}	
	\end{indented}
\end{table}
\begin{figure}
	\begin{indented}
	\item[]
	\includegraphics[width=5cm]{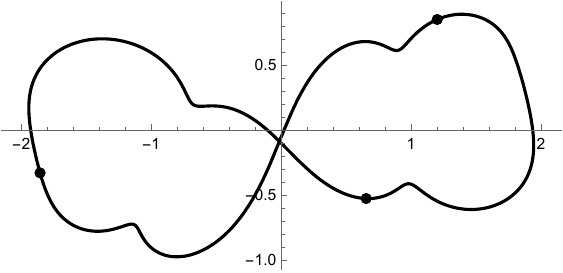}
	\caption{\label{fig:nonsc}
		Orbit at $T=45$ for non-symmetric choreography $q^{(b)^2}$ 
		with $C_6\{\mu{\cal C}\}C$  
		bifurcated at $T^{(b)^2}=19.020$ from $q^{(b)}$ with 
		$D_{6}\{\mu{\cal C},{\cal S}\}C_i$ 
		in table \ref{LJ18615}.	
		Initial conditions: 
		$(x_1,y_1) = (1.2013307,0.85317488)$, 
		$(x_2,y_2) = (-1.8559322,-0.32791656)$, 
		$(\dot{x}_1,\dot{y}_1) = (-0.30882003,-0.13297343)$ and 
		$(\dot{x}_2,\dot{y}_2) = (-0.041060813,0.14612712)$. 
	}
	\end{indented}
\end{figure}
\begin{table}
	\centering
	\caption{\label{LJ17132}
		$q^{(b)^2}$ from $q^{(b)}$ with $C_{6h}\{{\cal CM},\mu\}C_y$  
		from $\alpha^+$ solution $q$ 
		in table \ref{LJ}.
	}
	\begin{indented}
		\item[]	\begin{tabular}{cccclc}
			\hline
			$T^{(b)^2}$&branch& $\kappa$ & $D$& $G(q^{(b)^2})$  
			& side\\ 
			\hline
			$17.132$&$0$&$\nearrow$&$C_1$&$C_{6h}\{{\cal CM},\mu\} C_y$ 
			&R \\ 		 
			$17.235$&$+$&$\nearrow$&$C_3$&$D_{2}\{{\cal M},\mu\} D_y$  
			&B \\ 		 
			$17.785$&$+$&$\nearrow$&$C_6$&$C_{2}\{\mu\}D$ $C_{2}\{\mu\}D$ 
			& R\\
			\hline
		\end{tabular}	
	\end{indented}
\end{table}
\begin{figure}
	\begin{indented}
	\item[]
	\includegraphics[width=4.2cm]{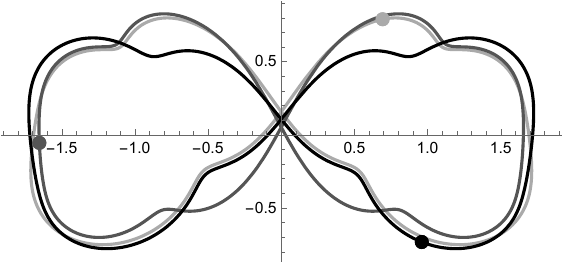}
	\includegraphics[width=4.2cm]{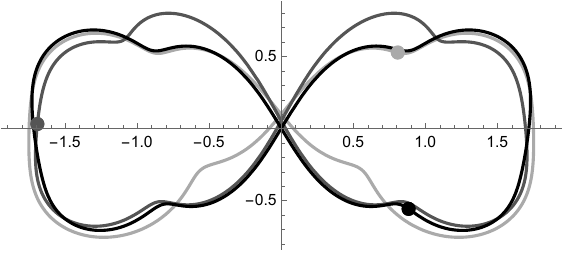}
	\includegraphics[width=4.2cm]{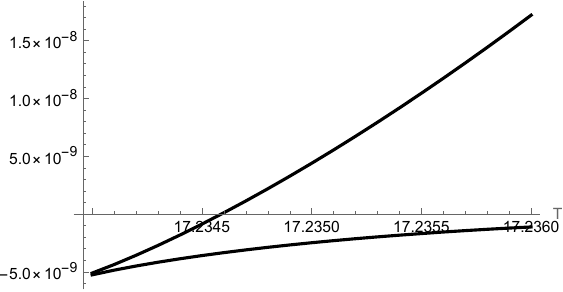}
	\\
	\footnotesize
	\hspace*{5em} (a) \hspace{10em} (b) \hspace{12em} (c)
	\caption{\label{fig:Dy}
		Orbits at $T=30$ for $y$-axis symmetric solution $q^{(b)^2}$ 
		with $D_2\{{\cal M}, \mu\}D_y$  
		\cite{fukuda2023} 
		and their action values, bifurcated at $T^{(b)^2}=17.235$ from $q^{(b)}$ 
		with $C_{6h}\{{\cal CM},\mu\}C_y$.  
		(a) $q^{(b)^2}$ from the left side of bifurcation point, 
		(b) from right side, 
		and
		(c) action value $S(q^{(b)^2})-S(q^{(b)})$.
		Initial conditions: (a) $(x_1, y_1) = (0, 0.10529629)$,
		$(x_2, y_2) = (1.5421890, 0.45426224)$,
		$(\dot{x}_1, \dot{y}_1) = (0.30563750, -0.37778377)$ and
		$(\dot{x}_2, \dot{y}_2) = (-0.17440457, 0.19744777)$.  
		(b) 
		$(x_1, y_1) = (0, 0.0014326920)$,
		$(x_2, y_2) = (1.6469521, 0.47710996)$,
		$(\dot{x}_1, \dot{y}_1) = (0.23166197, -0.38278558)$ and
		$(\dot{x}_2, \dot{y}_2) = (-0.12592133, 0.18861645)$. 
	}
	\end{indented}
\end{figure}
\begin{figure}
	\begin{indented}
		\item[]
		\includegraphics[width=4.2cm]{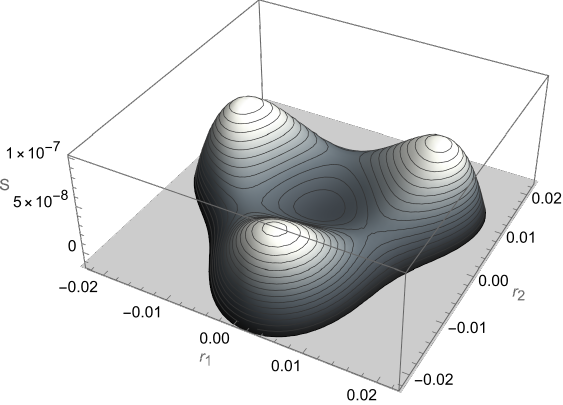}
		\caption{\label{fig:DyS}
			$S(q^{(b)}+\phi r)-S(q^{(b)})$, $r=(r_1,r_2)$ at $T=17.259$ 
			for $y$-axis symmetric solution $q^{(b)^2}$ 
			with $D_2\{{\cal M}, \mu\}D_y$  
			\cite{fukuda2023}  
			bifurcated at $T^{(b)^2}=17.235$ from $q^{(b)}$ 
			with $C_{6h}\{{\cal CM},\mu\} C_y$. 
		}
	\end{indented}
\end{figure}

\begin{figure}
	\begin{indented}
	\item[]
	\includegraphics[width=4.2cm]{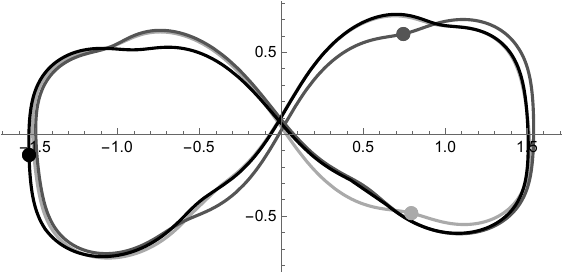}
	\includegraphics[width=4.2cm]{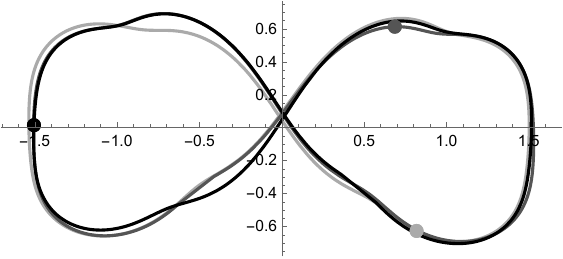}
	\includegraphics[width=4.2cm]{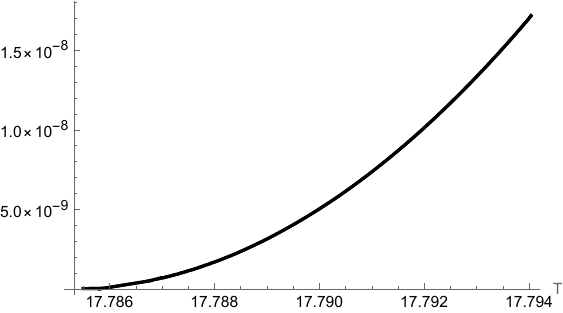}
	\\
	\footnotesize
	\hspace*{5em} (a) \hspace{10em} (b) \hspace{12em} (c)
	\caption{\label{fig:D1}
		Orbits at $T=20$ for non-symmetric solution $q^{(b)^2}$
		with $C_2\{\mu\}D$ and their action values, bifurcated at $T^{(b)^2}=17.785$ 
		from the $q^{(b)}$ with $C_{6h}\{{\cal CM},\mu\}C_y$. 
		(a) $q^{(b)^2}$ with higher action value, 
		(b) with lower action value, and
		(c) their action values $S(q^{(b)^2})-S(q^{(b)})$. 
		Initial conditions: (a) $(x_1, y_1) = (0, 0.045)$,
$(x_2, y_2) = (1.4396475, 0.5347018)$,
$(\dot{x}_1,\dot{y}_1)=(0.34271348, 0.30563750)$ and
$(\dot{x}_2,\dot{y}_2)=(-0.17598165, 0.25858920)$. 
(b) 		
$(x_1, y_1) = (0, 0.045)$,
$(x_2,y_2)=(1.4327086, 0.40489872)$,
$(\dot{x}_1,\dot{y}_1)=(0.35818585, -0.52798612)$ and
$(\dot{x}_2,\dot{y}_2)=(-0.12592133, 0.18861645)$. 
	}
\end{indented}
\end{figure}

In tables \ref{LJ18615} and \ref{LJ17132}, 
bifurcation solution $q^{(b)^2}$ 
at $T^{(b)^2}=(T^{(b)})^{(b)}$
from $q^{(b)}$ at 
$T^{(b)}=18.615$, $17.132$
in table \ref{LJ} 
are tabulated, respectively, in the same manner.
Since $q^{(b)}$ is bifurcated 
by one side bifurcation from $q$, 
$q^{(b)}$ begins at the bifurcation point $T^{(b)}$. 

Three planar solutions with symmetry 
$G(q^{(b)^2})=C_6\{\mu{\cal C}\}C$, $D_2\{{\cal M},\mu\}D_y$ and $C_2\{\mu\}D$, 
in tables \ref{LJ18615} and \ref{LJ17132},
are new types of solutions in our numerical calculations. 

\subsubsection{Non-symmetric choreography}
The two-fold type bifurcation solution $q^{(b)^2}$ at $T^{(b)^2}=19.020$ 
in tables \ref{LJ18615}
with symmetry $G(q^{(b)^2})=C_6\{\mu{\cal C}\}C$ 
is non-symmetric choreography. 
In figure \ref{fig:nonsc},  
its non-symmetric orbit at $T=45$ with initial conditions is shown.

\subsubsection{Three-fold type of $C_3$ representation}
The three-fold type bifurcation solution $q^{(b)^2}$ at $T^{(b)^2}=17.235$ 
in table \ref{LJ17132}
with symmetry $G(q^{(b)^2})=D_2\{{\cal M},\mu\}D_y$ 
has three separate $y$-axis symmetric orbits \cite{fukuda2023}. 
In figure \ref{fig:Dy},  
their orbits at $T=30$ with initial conditions 
and the action value $S(q^{(b)^2})-S(q^{(b)})$ around the  
bifurcation point are shown.
The solution to the left side of the bifurcation point 
soon turns back by fold bifurcation 
as another three-fold type \cite{fukuda2019}.   

This solution is the first numerical example with 
the three-fold type of $C_3$ representation 
where its $G(r;q^{(b)})$ does not depend on the direction of $r$, 
see $C_3\{C(3),+E\}$ in table \ref{D2h} (a). 

In figure \ref{fig:DyS}, three-dimensional  
plot $S(q^{(b)}+\phi r)-S(q^{(b)})$ against $r=(r_1,r_2)$
at $T=17.259$ is shown, which is the main term of $S(q^{(b)^2}(r;q))-S(q^{(b)})$.
The pond at $r=0=(0,0)$  
corresponds to the original solution $q^{(b)}$
since $\kappa>0$ in (\ref{S3}).   
Around it, three saddles correspond to the solutions bifurcated to 
the right side of the bifurcation point, and  
three peaks the solutions to the left side, turned back at $T=17.2340$.  
The saddles move to $r=0$ for $T \to 17.2346$, but the peaks do not. 
After three saddles coalesce for $T > 17.2346$ pond at $r=0$ turns to peak, 
and saddles move to the opposite side between the peak at $r=0$ and one of three peaks since $\kappa>0$.
This plot visually explains how we express the bifurcation solution 
by the variation principle of reduced action.

\subsubsection{Six-fold type of $C_6$ representation} 
The six-fold type bifurcation solutions $q^{(b)^2}$ at $T^{(b)^2}=17.785$ 
in table \ref{LJ17132} consist of two incongruent solutions
with the same symmetry $G(q^{(b)^2})=C_2\{\mu\}D$.  
In figure \ref{fig:D1} (a) and (b),  
their orbits at $T=20$  with initial conditions are shown.
The orbits of two solutions, (a) and (b), 
are very similar but not congruent and distinguishable.
Both orbits are eight-shaped, but have no spatial symmetry. 

In figure \ref{fig:D1} (c),  
their action values $S(q^{(b)^2})-S(q^{(b)})$ 
around the bifurcation point are shown. 
However, since the absolute difference 
between the two values is too small compared with $|S(q^{(b)^2})-S(q^{(b)})|$ 
it is impossible to distinguish the two curves in figure \ref{fig:D1} (c).

This solution is the first numerical example with 
the six-fold type of $C_6$ representation 
where its $G(r;q^{(b)})$ does not depend on the direction of $r$,  
and two kinds of solutions have the same symmetry, 
see $C_6\{C(6),+E\}$ in table \ref{D2h} (a).

\subsection{Homogeneous potential system}

\begin{table}
	\centering
	\caption{\label{H8}
		$q^{(b)}$ from $q$ 
		with $D_{6h}\{{\cal CM},{\cal S},\mu\}C_{xy}$  
		for $u(r)=-1/r^{a}$. 
	}
	\begin{indented}
	\item[]	
	\begin{tabular}{rccclc}
		\hline
		\multicolumn{1}{c}{$a^{(b)}$}&& $\kappa$ & $D$& $G(q^{(b)})$ 
		&side\\
		\hline
		$-0.3817$&&$\nearrow$& $D_6$
		& $D_{2}\{{\cal C}^{-n}{\cal SM},\mu\}D_x$  $D_{2}\{{\cal C}^{1-n}{\cal S},\mu\}D_i$ 
		&\\  
		$-0.2142$&&$\nearrow$& $C_2$& $D_{6}\{\mu{\cal C},{\cal SM}\} C_x$  
		&R\\  
		$0.9966$&&$\nearrow$& $D_3$& $D_{2h}\{{\cal C}^{n-1}{\cal S},{\cal M},\mu\} D_{xy}$ 
		&B
		\\  
		$1.3424$&&$\nearrow$& $D_6$
		& $D_2\{{\cal C}^{-n}{\cal SM},\mu\}D_x>D_2\{{\cal C}^{1-n}{\cal S},\mu\}D_i$  
		&R\\  
		\hline
	\end{tabular}	
	\end{indented}
\end{table}
In table \ref{H8}, bifurcations calculated numerically
for the system with the homogeneous potential 
\begin{equation}
	u(r)=-\frac{1}{r^a}
\end{equation}
from the figure-eight choreography $q$ \cite{fukuda2019,fujiwara2020}, 
are shown in the same manner. 
The bifurcation parameter $\xi$ is the power $a$ and 
the bifurcation point $\xi^{(b)}$ is $a^{(b)}$ tabulated in the first column. 
Note that bifurcation at $a^{(b)}=-0.3817$ is identified 
by eigenvalue $\kappa$ and eigenfunction $\phi$, 
but the bifurcation solution is not calculated numerically, 
and its `side' and inequality between symmetries are empty.

\begin{table}
	\centering
	\caption{\label{simoH}
		$q^{(b)^2}$ from $q^{(b)}$ 
		with $D_{2h}\{{\cal C}^{n-1}{\cal S},{\cal M},\mu\}D_{xy}$, 
		$n=1$, in table \ref{H8}.
	}
	\begin{indented}
	\item[]	
	\begin{tabular}{cccclc}
		\hline
		$a^{(b)^2}$&branch& $\kappa$ & $D$& $G(q^{(b)^2})$ 
		&side 
		\\ \hline
		$0.9100$&$-$&$\nwarrow$& $C_2$& $D_{2}\{{\cal S},\mu\} D_i$ 
		&R\\ 		 
		$0.8156$&$0$&$(\uparrow$&$C_1$&$D_{2h}\{{\cal S},{\cal M},\mu\} D_{xy}$ 
		&B\\ 
		$0.8460$&$+$&$\nearrow$& $C_2$ & $D_{2}\{\mu{\cal S},\mu{\cal M}\} 
		\bi{D}_{\hat{x}\hat{y}}$ 
		&L\\  
		$0.9966$&$0$&$\searrow$&$C_1$&$D_{2h}\{{\cal S},{\cal M},\mu\} D_{xy}$ 
		&X \\  
		$0.9966$&$0$&$\nearrow$&$C_2$&$D_{2}\{{\cal M},\mu\} D_y$ 
		&X \\  
		$1.0000$&$-$&\multicolumn{2}{c}{Sim\'{o}'s H}&$D_{2h}\{{\cal S},{\cal M},\mu\} D_{xy}$ 
		&  \\  
		$1.0272$&$0$&$\uparrow)$&$C_1$&$D_{2h}\{{\cal S},{\cal M},\mu\} D_{xy}$ 
		&B \\  
		\hline
	\end{tabular}	
	\end{indented}
\end{table}
\begin{figure}
	\begin{indented}
	\item[]
	\includegraphics[width=5cm]{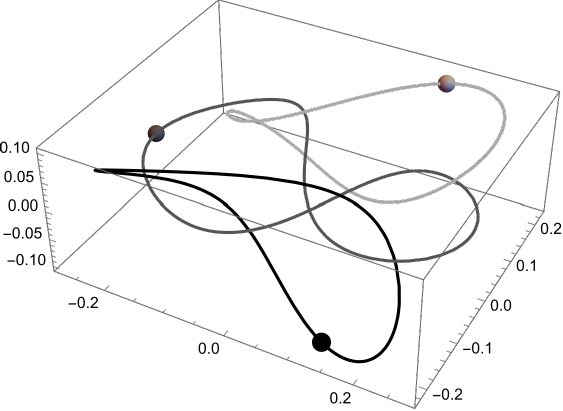}
	\caption{\label{fig:nonp}
		Orbit at $a=0.75$ for non-planar solution $q^{(b)^2}$ 
		with $D_{2}\{\mu{\cal S},\mu{\cal M}\}\bi{D}_{\hat{x}\hat{y}}$  
		bifurcated at $a^{(b)^2}=0.8460$ by two-fold type bifurcation 
		from $q^{(b)}$ with 
		$D_{2h}\{{\cal C}^{n-1}{\cal S},{\cal M},\mu\}D_{xy}$,  
		$n=1$ in table \ref{simoH} \cite{fukuda2020}.
		Initial conditions:     
		$(x_1,y_1,z_1)=(0.13030202, 0.19616588, 0.098259590)$, 
		$(x_2,y_2,z_2)=(-2x_1, 0, 0)$, 
		$(\dot{x}_1,\dot{y}_1,\dot{z}_1)=(-v\cos\theta, -v\sin\theta, -0.013911923)$, 
		$(\dot{x}_2,\dot{y}_2,\dot{z}_2)=(0, -2\dot{y}_1, -2 \dot{z}_1)$, 
		$v=1.0045200$, 
		$\theta=\arctan(y_1,3x_1)$ and
		$T=1$.		
	}
	\end{indented}
\end{figure}
In table \ref{simoH}, we tabulated the 
bifurcation solution $q^{(b)^2}$ at $a^{(b)^2}=(a^{(b)})^{(b)}$ from 
$q^{(b)}$ at $a^{(b)}=0.9966$ in table \ref{H8},   
which becomes Sim\'{o}'s H solution at $a=1$ \cite{simoH}.
The row at $a^{(b)^2}=1.0000$ is not bifurcation but its indication.  
At $a^{(b)^2}=0.9966$,  
two eigenvalues for one-dimensional representation 
denoted by `X' in the `side' column cross.
This is a point where the solutions $q^{(b)}$ bifurcate from $q$.
We will comment on this in section \ref{sec:sum}.

\subsubsection{Non-planar bifurcation solution}
The two-fold type solution $q^{(b)^2}$ bifurcated at $a^{(b)^2}=0.8460$ 
with symmetry 
$G(q^{(b)^2})=D_2\{\mu{\cal S},\mu{\cal M}\}\bi{D}_{\hat{x}\hat{y}}$
in table \ref{simoH} 
is only non-planar solution among our numerical calculations \cite{fukuda2020}, 
whereas there is a non-planar solution by taking mass 
as a bifurcation parameter \cite{doedel2003}.
Its orbit at $a=0.75$ is shown in figure \ref{fig:nonp} with initial conditions.

\section{Summary and discussions}\label{sec:sum}

We could explain all bifurcations $q^{(b)^k}$ numerically observed from
the figure-eight choreography $q$ by the irreducible representation $D(q^{(b)^{k-1}})$ representing eigenspace of $H(q^{(b)^{k-1}})$.
They appeared to be the following four types of bifurcation solutions. 
Type 1: the trivial type bifurcates two solutions 
on both sides of the bifurcation point in $\kappa$, 
sometimes in the fold-bifurcation in $\xi$. 
Type 2: The two-fold type bifurcates two congruent solutions on one side. 
Type 3: The three-fold type bifurcates two kinds of three congruent solutions 
on both sides. 
Type 4: The six-fold type bifurcates two kinds of six congruent solutions on
one side.

For each bifurcation type,  
there is a condition of the coefficients, 
$A_3 \ne 0$ for type 1, $A_4 \ne 0$ for type 2, 
$A_3' \ne 0$ for type 3 and $A_4^{(0)} A_6' \ne 0$ for type 4.
Though we have not checked them for the numerical solutions, 
we expected from their symmetries they would belong to the four types. 

Accordingly, we almost understand how to yield the bifurcation solution $q^{(b)}$ 
from the original solution $q$. However, we have 
not described the inverse process, how to yield $q$ from $q^{(b)}$ yet.
From the numerical calculation,  
we observe that the inverse process will be represented 
by the eigenfunctions of $H(q^{(b)})$ 
with the same number and the same symmetry as the original process 
described by $H(q)$. 

\begin{figure}
	\begin{indented}
	\item[]
	\includegraphics[width=5cm]{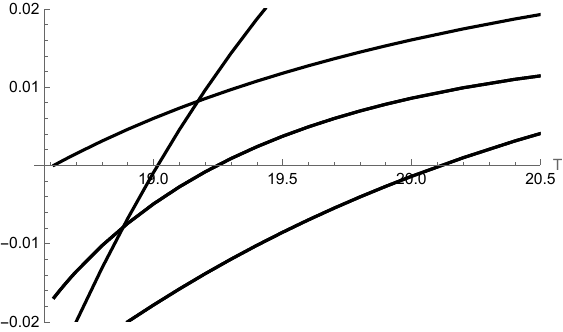}
	\hspace{1em}
	\includegraphics[width=5cm]{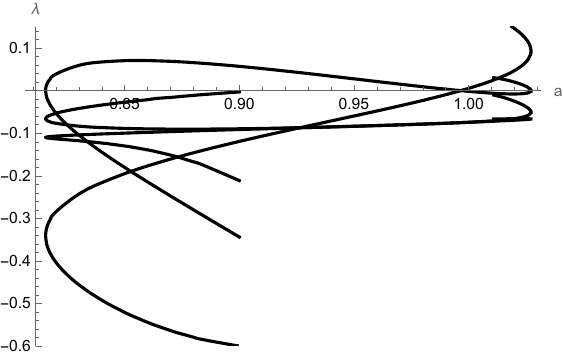}
	\\
	\footnotesize
	\hspace*{7em} (a) \hspace{14em} (b)
	\caption{\label{fig:lambda18615}
		Eigenvalues of $H(q^{(b)})$ 
		(a) for $G(q^{(b)})=D_{6}\{\mu{\cal C},{\cal S}\}D_i$
		\cite{fukuda2023} and
		(b) for planar eigenfunctions for $G(q^{(b)})=D_{2h}\{{\cal S},{\cal M},\mu\}D_{xy}$. 
	}
	\end{indented}
\end{figure}
In figure \ref{fig:lambda18615} (a), 
eigenvalues of $H(q^{(b)})$ 
for $G(q^{(b)})=D_{6}\{\mu{\cal C},{\cal S}\}D_i$ 
in table \ref{D6} are shown \cite{fukuda2023}.
The eigenfunction of the curve 
going to zero toward $T=18.615$ has the same symmetry as  
$q^{(b)}$ and represents the inverse process.
The original bifurcation is represented by a one-dimensional eigenfunction 
with the same $D_{6}\{\mu{\cal C},{\cal S}\}D_i$ symmetry.

In figure \ref{fig:lambda18615} (b), 
planar eigenvalues for $G(q^{(b)})=D_{2h}\{{\cal S},{\cal M},\mu\}D_{xy}$  
in table \ref{simoH}  
are shown \cite{fukuda2019p}.
The eigenfunctions of the non-degenerate two curves crossing at $a=0.9966$ have 
the same symmetries with the eigenfunctions representing the original process, 
$D_{2h}\{{\cal C}^{n-1}{\cal S},{\cal M},\mu\}D_{xy}$, $n=1$,  
and $D_{2}\{{\cal M},\mu\}D_y$  
of doubly degenerate eigenvalue in table \ref{D6h}.
As in the one-dimensional case in figure \ref{fig:lambda18615} (a), 
one of the eigenfunctions has the same symmetry as $q^{(b)}$.

Further, we assumed implicitly that the  
symmetry group commuting with $H(q)$ does not change. 
Thus, our method of analysis is not applicable to the bifurcation by changing the masses, where the symmetry of $H(q)$ changes.
In future work, the symmetry changing system 
and the above-mentioned inverse bifurcation process $q^{(b)} \to q$  
should be included in the theory.  

In numerical search, 
we found a non-planar bifurcation solution, 
which is unfortunately not choreographic, 
in the homogeneous interaction potential system.
Though the LJ potential system possesses a lot of bifurcations,  
we could not find any non-planar solution. 
We do not know the reason why the remarkable 
`non-planar figure-eight choreographic' bifurcation solution, 
up to now theoretically possible, has not been found.

Our method of analysis will be applicable to 
the bifurcation of the periodic solution for general few body systems 
if it has symmetries. 

\ack
We wish to thank Professor Toshiaki Fujiwara for many useful discussions.

\end{document}